\documentclass[conference,10pt]{IEEEtran}
\usepackage{epsfig,makeidx,color}
\usepackage{amsmath,amssymb,bbm}
\usepackage{cite,graphicx,lipsum}
\usepackage{enumerate}
\usepackage[switch,pagewise]{lineno}
\usepackage{hyperref}
\hypersetup{
        colorlinks = true,
        citecolor=blue,
}
\def\cK{{\cal K}}

\def\uP{{\mathbb P}}

\def\uE{{\mathbb E}}

\newtheorem{mytheorem}{\bf Theorem} 
\newtheorem{mylemma}{\bf Lemma} 

\def\be{ \begin{equation} }
\def\ee{ \end{equation} }
\def\bea{ \begin{eqnarray} }
\def\eea{ \end{eqnarray} }

\def\by{{\bf y}}
\def\bc{{\bf c}}

\def\bs{{\bf s}}

\def\bn{{\bf n}}

\def\bC{{\bf C}}

\def\bI{{\bf I}}

\def\b0{{\bf 0}}

\def\cC{{\cal C}}

\def\cN{{\cal N}}
\def\cS{{\cal S}}

\ifCLASSOPTIONonecolumn
  \newcommand{\figwidth}{0.55\columnwidth}
  
\else
  \newcommand{\figwidth}{0.80\columnwidth}
  
\fi

\begin{document}

\title{Multichannel ALOHA with Exploration Phase}

\author{\IEEEauthorblockN{Jinho Choi}
\IEEEauthorblockA{School of Information Technology,
Deakin University \\
Geelong, VIC 3220, Australia \\
e-mail: jinho.choi@deakin.edu.au}
}

%

\maketitle
\begin{abstract}
In this paper, we consider exploration for multichannel 
ALOHA by transmitting preambles before transmitting data packets
and show that the maximum throughput
can be improved by a factor of $2 - e^{-1} \approx 1.632$,
which can be seen as the gain of exploration.
In the proposed approach,
a base station (BS) needs to send the feedback information 
to active users to inform
the numbers of transmitted preambles in multiple channels,
which can be reliably estimated as in compressive random access.
Simulation results also confirm the results from analysis.
\end{abstract}


\ifCLASSOPTIONonecolumn
\baselineskip 28pt
\fi

\section{Introduction}

In order to support the connectivity of a large number
of devices and sensors for the Internet of Things (IoT),
machine-type communication (MTC)
has been considered in cellular systems
\cite{Taleb12} \cite{3GPP_MTC} \cite{3GPP_NBIoT}.
In fifth generation (5G) systems,
it is expected to have more standards for
MTC \cite{Condoluci15} \cite{Shar15} \cite{Bockelmann16}.
In general, in order to support a large number of devices
(in this paper, we assume that devices and users are 
interchangeable)
with sparse activity, i.e., only a fraction of them 
are active at a time, random access is widely considered 
as it can avoid high signaling overhead.
In particular,
most MTC schemes are based on (slotted) ALOHA \cite{Abramson70},
and ALOHA is extensively studied for MTC as in
\cite{Arouk14} \cite{Lin14} \cite{Chang15} \cite{Choi16}.

Since ALOHA plays a key role in MTC,
various approaches are considered for ALOHA 
in order to improve the performance 
in terms of throughput
(which may result in the increase of the number of devices
to be supported). In \cite{Casini07},
contention resolution repetition diversity (CRRD) 
is considered 
together with successive interference cancellation (SIC)
for a better throughput. 
The notion of coding is applied to CRRD in \cite{Liva11}
\cite{Paolini15},
where it is shown that the throughput 
(in the average number of successfully transmitted packets per slot)
can approach 1.

In \cite{Choi_JSAC} \cite{Choi18b},
the notion of non-orthogonal multiple access (NOMA)
\cite{Choi08}
is applied to ALOHA so that multiple virtual access channels
are created in the power domain without any bandwidth expansion,
and it is shown that the throughput can be significantly
improved at the cost of high power budget at users.
In \cite{Seo18}, the performance of NOMA-based random access
is further analyzed.

Provided that a wide bandwidth is available (e.g., the bandwidth
is $B$),
there can be multiple ALOHA systems (of a bandwidth of $B/M$,
where $M$ is the number of systems or channels)
which results in multichannel ALOHA \cite{Shen03}
\cite{Arouk14} \cite{Chang15}.
It is shown that the throughput 
grows linearly with the number of 
channels. Thus, the total throughput per Hz 
becomes independent
of the number of channels for a fixed bandwidth,
which means that there is no advantage 
of multichannel ALOHA over single-channel
ALOHA of wideband in terms of throughput.

In this paper, we consider 
an exploration approach for multichannel ALOHA to improve
the performance.
As in multi-armed bandit problems
\cite{Bianchi06} \cite{Cohen07},
exploration can help improve the performance.
For exploration, in the proposed approach,
each active user, i.e., a user
with packet to transmit, is to send a preamble
prior to packet transmission,
and a base station (BS) sends the feedback information 
to active users to inform
the numbers of transmitted preambles in multiple channels.
We show that the feedback information,
which is the outcome of exploration,
can improve the throughput of multichannel ALOHA
(as well as single-channel ALOHA).
In particular, in terms of the maximum throughput,
it is shown that the performance can be improved by a factor
of $2 -e^{-1} \approx 1.632$ thanks to exploration.

It is noteworthy that the exploration by sending preambles
becomes possible if the BS is able to estimate
the number of transmitted preambles in each channel.
Thanks to the notion of compressive random access 
\cite{Applebaum12}  \cite{Choi_CRA18},
the BS can estimate the number of 
transmitted preambles in each channel precisely.
In addition, the proposed approach
does not use SIC and CRRD, which makes it easy to implement.

In summary, the main contributions of the paper
are as follows: 
\emph{i)} a multichannel ALOHA scheme is proposed
with exploration to improve the throughput;
\emph{ii)} performance analysis is carried out,
which shows that the maximum throughput of the proposed
is higher than that of conventional ALOHA by a factor of
$2 - e^{-1}$.


\section{Motivation and System Model}	\label{S:SM}

Throughout this paper, we only consider a slotted ALOHA
system consisting of one BS and multiple users for uplink
transmissions,
where the BS periodically transmits a beacon signal 
for synchronization.

\subsection{Motivation}

Consider a single-channel ALOHA system.
Let $T_{\rm d}$ denote the length of data packet.
If an active user (with a data
packet to transmit) knows that there are other active
users, she may not transmit to avoid collision.
In order to see whether or not there are 
other active users, suppose that
each active user can transmit a preamble
sequence before data packet transmission, 
which can be seen as the {\it exploration}
to learn the environment.
Let $T_{\rm p}$ denote the length of preamble.
It is assumed that $T_{\rm p} < T_{\rm d}$ in general.
At the end of preamble transmission, 
we assume that the BS is able to detect
all the transmitted preamble sequences
and sends a feedback signal to inform
the number of the transmitted preamble sequences.
The length of feedback signal is denoted by $T_{\rm f}$.

An active user can make a decision whether or not
she transmits her data packet based on the feedback from the BS.
If there is only one preamble transmitted, 
the active user should send a data packet as there is no
other active user. However, if the number of 
transmitted preambles is larger than $1$, 
each active user may transmit 
a packet with a certain probability that might be
less than 1.
For example, in order to maximize the probability of successful
transmission, 
the access probability might be $\frac{1}{K}$,
where $K$ represents the number of active users
or transmitted preambles that is fed back from the BS.
Therefore, for a given $K \ge 1$, the throughput,
which is the average number of transmitted packets without collisions,
becomes
\be
\eta_{\rm sa} (K) = \left( 1 - \frac{1}{K} \right)^{K-1} \ge e^{-1}.
	\label{EQ:ineq}
\ee
As $K \to \infty$, we can see that 
$\eta_{\rm sa}$ approaches $e^{-1}$.
On the other hand, if $K = 1$, $\eta_{\rm sa} = 1$.
From \eqref{EQ:ineq}, the average throughput can be shown to be
higher than $e^{-1}$ as follows:
\be
\uE[\eta_{\rm sa} (K)] \ge e^{-1}.
	\label{EQ:esa}
\ee

\begin{figure}[thb]
\begin{center}
\includegraphics[width=\figwidth]{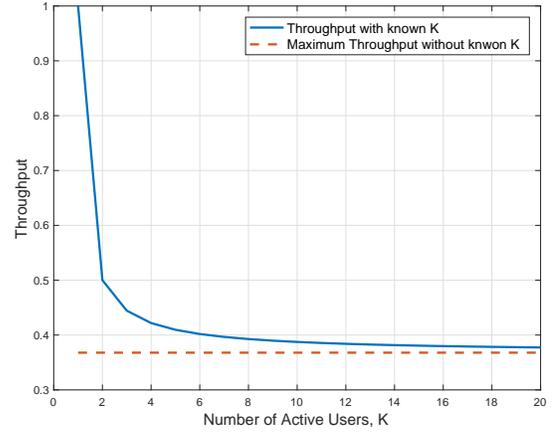}
\end{center}
\caption{Throughout with known number of active users, $K$,
and the maximum throughput without knowing $K$, i.e., $e^{-1}$.}
        \label{Fig:gap_sa}
\end{figure}

On the other hand, suppose that
the access probability, denoted by $p$, is decided without knowing $K$.
In this case, if $K$ is a Poisson random variable
with mean $\lambda$, 
where $\lambda$ is seen as the packet arrival rate,
the throughput becomes
\be
\eta_{\rm sa}  = p \lambda e^{- p \lambda} \le e^{-1},
	\label{EQ:esa_p}
\ee
where the upper-bound can be achieved by $p = \frac{1}{\lambda}$ for
$\lambda \ge 1$.
Therefore, there is a gain\footnote{The gain can be offset by
the overhead due to exploration, i.e., the overhead due
to preamble transmissions.
However, if $T_{\rm p} \ll T_{\rm d}$, the offset might be
negligible. We will discuss more details later.} 
(i.e., the difference between \eqref{EQ:esa} and \eqref{EQ:esa_p})
obtained by the exploration that
allows active users to know how many are in contention,
i.e, the number of active users, $K$.
In addition, as shown in 
\eqref{EQ:ineq}, the gain increases if $K$ is small,
which is also illustrated in Fig.~\ref{Fig:gap_sa}.

\subsection{System Model}

In this subsection,
we generalize the exploration in the previous subsection
to multichannel ALOHA.

Suppose that there are $M$ orthogonal
radio resource blocks for multiple access channels.
We assume that each active user
can randomly choose one channel and
transmit a preamble signal to the BS
in the exploration phase (EP).
After the EP, the BS can find 
the number of active users for each channel.
Let $k_m$ denote 
the number of active users transmitting their preambles
through the $m$th channel.
Then, the BS broadcasts the numbers of active
users for all the channels, $\{k_1, \ldots, k_M\}$
so that all the active users can see the state of
contention.
For example, an active user transmits a preamble
through the 1st channel and sees that $k_1 = 1$.
In this case, clearly,
the user is only one active user using 
channel 1.

Let 
$\cS = \{m \,|\, k_m = 1, \ m = 1,\ldots, M\}$,
i.e., $\cS$ is the index set of the channels with only
one active user transmitting preamble.
For convenience, let $\cS^c$ denote
the complement of $\cS$.
In the data transmission phase (DTP),
if a user transmitting a preamble through 
channel $m$ sees that $k_m = 1$ or $m \in \cS$, the user
can send a data packet through the $m$th channel.
For convenience, this user is referred to as a contention-free
user. The group of contention-free active users 
is also referred to as Group I.
On the other hand, when $m \in \cS^c$ (which implies
that $k_m \ge 2$ as the user transmits a preamble
through channel $m$ and $m \notin \cS$),
the user is referred to as a user in contention, and the
group of active users in  contention is referred to as
Group II.

\begin{figure}[thb]
\begin{center}
\includegraphics[width=\figwidth]{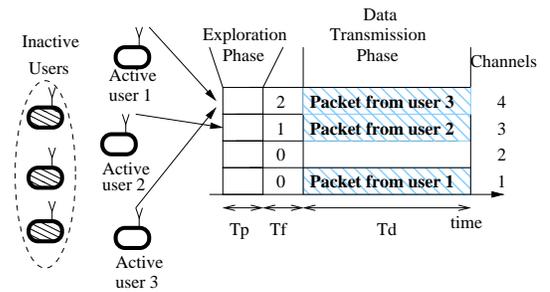}
\end{center}
\caption{An example
with $K = 3$ and $M = 4$, where active user 2 chooses
channel 3 to transmit a preamble and two other
active users (user 1 and 3) choose channel 4.}
        \label{Fig:sys}
\end{figure}

An example is shown in Fig.~\ref{Fig:sys}
with $K = 3$ and $M = 4$. Active user 2 chooses
channel 3 to transmit a preamble and two other
active users (user 1 and 3) choose channel 4.
From this, the feedback information from the BS to the
active users is $\{k_1, k_2, k_3, k_4\}
= \{0, 0, 1, 2\}$.
As a result, $\cS = \{3\}$ and $\cS^c = \{1, 2, 4\}$,
and active user 2 belongs to Group I and 
active users 1 and 3 belong to Group II.

If the user in contention transmits a packet through channel
$m$, it will be collided with others. 
To avoid packet collision, 
the user can choose a different channel, say channel $l
\in \cS^c$, and
transmits a packet. However, since another user in contention
can choose channel $l$, there can be packet collision.

Although it is not possible to prevent collisions for users
in contention, in order to mitigate packet collisions, we can assume that
each user in contention can randomly choose a channel in $\cS^c$
with an access probability, $p_{\rm dtp}$, in the DTP
to transmit a packet, while
a user in contention does not transmit a packet with
probability $1 - p_{\rm dtp}$.

\section{Maximum Throughput Comparison}	\label{S:MT}

In this section, we focus on the maximum throughput
for two schemes, namely conventional multichannel ALOHA
and multichannel ALOHA with EP,
and show that the ratio of the maximum
throughput of 
multichannel ALOHA with EP to that of
conventional multichannel ALOHA becomes $2 - e^{-1} \approx 1.632$,
which can be seen as the gain of exploration.

\subsection{Throughput of Conventional Multichannel ALOHA}

In conventional multichannel ALOHA,
we can find the 
average number of packets without collisions
as follows:
\be
N_{\rm ma} (K,M) = K \left(1 - \frac{1}{M} \right)^{K-1}.
\ee
For a large $K$, it can be shown that
\begin{align}
N_{\rm ma} (K,M)
& = K \left(1 - \frac{1}{M} \right)^{K-1} \cr
& \approx K e^{- \frac{K}{M}}  \le \bar N_{\rm ma} (M) = M e^{-1},
	\label{EQ:ma}
\end{align}
where $\bar N_{\rm ma}(M)$ is the maximum throughput
of multichannel ALOHA,
which can be achieved if $K = M$.
In addition, as in \cite{Shen03},
the arrival rate has to be lower than $M e^{-1}$ for 
system stability.

\subsection{Throughput of Multichannel ALOHA with EP}

One of the main results 
of the paper can be stated as follows.

\begin{mytheorem}	\label{T:1}
The ratio of the maximum throughput of  multichannel ALOHA with EP 
to that of conventional multichannel ALOHA is
\be
\eta = 2 - e^{-1}.
	\label{EQ:T1}
\ee
\end{mytheorem}
In the rest of this subsection, we prove Theorem~\ref{T:1}.

Let $S = |\cS|$,
where $S$ represents the number of active users
that can transmit packets without collisions.
Clearly, $S \le \min\{K, M\}$.
In addition, let $W = K - S$, where
$W$ becomes the number of active users in contention.
For convenience, let
$ L = M - S$.
Clearly, $L$ is the number of the channels that are available
for contention-based transmissions for $W$ active users in contention
or Group II.
In addition, suppose that among $W$,
$U$ active users in contention are to transmit 
their packets through $L$ channels.
Clearly, for given $W$, $U$ has the following  
distribution:
\be
\uP(U = u\,|\,W) = \binom{W}{u} p_{\rm dtp}^u (1 - p_{\rm dtp})^{W-u}.
	\label{EQ:Pu}
\ee
Throughout the paper, we assume that
\be
p_{\rm dtp} = \min \left\{
1, \frac{L}{W} \right\},
	\label{EQ:p_dtp}
\ee
which maximizes\footnote{For Group II, we can consider
a multichannel ALOHA system with $L$ channels and $W$ users. 
In this case, $p_{\rm dtp}$ is seen
as the access probability that can maximize
the throughput if it is given in \eqref{EQ:p_dtp}
\cite{Shen03}, \cite{Chang15}.}
the average number of 
packets without collisions in Group II.

For a given $U$,
the conditional average number of the packets that can be successfully
transmitted from $U$ users without collisions in contention during DTP
becomes $U \left(1 - \frac{1}{L} \right)^{U-1}$.
As a result, the 
conditional average number of packets without collisions
(for given $U$, $L$, and $S$)
is given by
\be
N_{\rm ep} (U, L, S) =  S + U \left(1 - \frac{1}{L} \right)^{U-1},
	\label{EQ:VULS}
\ee
where the first term on the right-hand side (RHS) in \eqref{EQ:VULS}
is the number of packets without collisions from Group I
and the second term is that from Group II.
For convenience, let 
\be
N_{\rm ep} (K,M) = \uE[N_{\rm ep} (U, L, S) \,|\, K],
\ee
which is the average number of 
packets without collisions (for given $K$ and $M$) in multichannel ALOHA
with EP.

\begin{mylemma}
Suppose that $M$ and $K$ are sufficiently large
so that $L$ and $W$ are also large.
The upper bound on 
$N_{\rm ep} (K,M)$ is given by
\be
N_{\rm ep} (K,M)  \le M e^{-1} + \bar S (K) (1 - e^{-1}).
	\label{EQ:L1}
\ee
\end{mylemma}
\begin{IEEEproof}
From Eq.~\eqref{EQ:VULS}, we have
\begin{align}
N_{\rm ep} (K,M) & = \uE[ N_{\rm ep} (U, L, S) \,|\, K] \cr
& =  \uE[ S\,|\,K] + 
\uE\left[U \left(1 - \frac{1}{L} \right)^{U-1}\,|\, K
\right].
	\label{EQ:Vep1}
\end{align}
In \eqref{EQ:Vep1}, it can be shown that
\begin{align}
\bar S(K) = \uE[S \,|\, K] 
= K \left(1 - \frac{1}{M} \right)^{K-1}.
	\label{EQ:bSK}
\end{align}
Since $U$ depends on $W$ and $W$ depends on $K$,
we have
$$
\uE \left[  U \left(1 - \frac{1}{L} \right)^{U-1}
 \,\bigl|\,  K \right]
=
\uE \left[ \left[ U \left(1 - \frac{1}{L} \right)^{U-1}
 \,\bigl|\, W \right]\,\bigl|\, K \right]
$$
From \eqref{EQ:Pu},
after some manipulations,
it can be shown that
\begin{align}
\uE\left[ U \left(1 - \frac{1}{L} \right)^{U-1}
 \,|\, W \right] 
& = \sum_{u=0}^W \uP(U = u \,|\, W) 
u \left(1 - \frac{1}{L} \right)^{u-1} \cr
& = p_{\rm dtp} W \left(
1 - \frac{p_{\rm dtp}}{L} \right)^{W-1}.
\end{align}
If $\frac{p_{\rm dtp}}{L} \ll 1$,
it follows that
$$
p_{\rm dtp} W \left(
1 - \frac{p_{\rm dtp}}{L} \right)^{W-1}
\approx 
p_{\rm dtp} W e^{-\frac{p_{\rm dtp} W}{L}}
\le L e^{-1}.
$$
The upper bound can be achieved if 
\be
p_{\rm dtp} = \frac{L}{W} \le 1.
	\label{EQ:op}
\ee
As a result, since $L = M - S$, it can be shown that
\begin{align}
\uE \left[  U \left(1 - \frac{1}{L} \right)^{U-1}
 \,\bigl|\,  K \right]
& \le \uE[L e^{-1}\,|\, K] \cr
& = (M - \uE[S\,|\, K])e^{-1}.
	\label{EQ:EU}
\end{align}
Substituting \eqref{EQ:bSK} and \eqref{EQ:EU}
into \eqref{EQ:Vep1}, we have
\begin{align}
N_{\rm ep} (K,M) & \le \bar S(K) +(M - \bar S(K) ) e^{-1} \cr
& = M e^{-1} + \bar S(K) (1 - e^{-1}), 
\end{align}
which completes the proof.
\end{IEEEproof}

From \eqref{EQ:L1},
the maximum throughput of multichannel ALOHA with EP 
is given by 
\begin{align}
\bar N_{\rm ep} (M) & = \max_K N_{\rm ep} (M,K) \cr
& = M e^{-1} + (1- e^{-1}) \max_K \bar S(K).
\end{align}
For a sufficiently large $K$, $\bar S(K) \approx K e^{-\frac{K}{M}}$. 
Thus, if we consider the throughput gain
using the ratio of the maximum throughput of multichannel ALOHA
with EP to that of conventional ALOHA, it can be shown that
\begin{align}
\eta & = \frac{\bar N_{\rm ep} (M)}{\bar N_{\rm ma} (M)} 
= \frac{M e^{-1} + (1- e^{-1}) M e^{-1}}{M e^{-1}} \cr
& = 2 - e^{-1} \approx 1.632,
	\label{EQ:eta}
\end{align}
which finally proves Theorem~\ref{T:1}.
From this, it is clear that
the EP can improve the performance
of multichannel ALOHA (in terms of the throughput) by a factor of 1.632.

\section{Implementation Issues}	\label{S:Imp}

In this section, we discuss a few key issues in implementations
including the cost for exploration and the estimation
of the number of transmitted preambles.

\subsection{Cost for Exploration}

In multichannel ALOHA with EP, 
each active user has to transmit 
a preamble prior to packet transmission. 
Thus, the length of slot in multichannel ALOHA with EP is 
$T_{\rm p}+ T_{\rm d} + 2 T_{\rm f}$,
while that in 
conventional ALOHA is $T_{\rm d} + T_{\rm f}$.
In  multichannel ALOHA with EP,
there are two types of feedback: one is for the numbers of transmitted
preambles in $M$ channels and the other 
is for the collisions of packets from the active 
users in Group II.
Thus, the following factor can be considered:
\be
\kappa = 
\frac{T_{\rm d} + T_{\rm f}}
{T_{\rm p}+ T_{\rm d} + 2 T_{\rm f}} = \frac{1}{1 + \epsilon_T},
\ee
where $\epsilon_T = \frac{T_{\rm p} + T_{\rm f}}{T_{\rm d} + T_{\rm f}}$.
The effective throughput of 
multichannel ALOHA with EP for
the comparison with that of conventional
multichannel ALOHA becomes $\kappa N_{\rm ep} (M)$.
As a result, if 
$$
\kappa N_{\rm ep} (M) > N_{\rm ma} (M),
$$
the exploration 
becomes beneficial to improve the performance
of multichannel ALOHA.
As mentioned earlier, 
since $T_{\rm d} \gg T_{\rm p}$, we can see that
$\kappa \approx 1$. Thus, in general, 
multichannel ALOHA with EP can have a better performance than
conventional multichannel ALOHA.

\subsection{Determination of Number of Transmitted Preambles}

In multichannel ALOHA with EP, 
we can consider two different cases when 
an active user transmits a preamble. 
In the first case, it is assumed that there is a common
set or pool of preambles,
denoted by $\cC =\{\bc_1, \ldots, \bc_{L}\}$. 
Here, $\bc_l$ represents the $l$th preamble sequence
(of length $T_{\rm p}$) and $L$ denotes the number of the preambles
in $\cC$. Any active user
is to randomly choose one in $\cC$.
At the BS, the signal received through the $m$th channel can be
expressed as follows:
\be
\by_m = \bC \bs_m + \bn_m,
\ee
where $\bC = [\bc_1 \ \ldots \ \bc_L]$ and $\bn_m \sim \cC \cN(\b0,
N_0 \bI)$ is the background noise vector.
Here, the $l$th element of $\bs_m$, denoted by $s_{m,l}$, is given by
$s_{m,l} = \sum_{k \in \cK_{m,l}} h_k \sqrt{P_k}$,
where $h_k$ is the channel coefficient from active user $k$
to the BS, $P_k$ is the transmit power of active user $k$,
and $\cK_{m,l}$ is the index set of the active users choosing the 
$l$th preamble in the $m$th channel. 
Suppose that active users can decide their transmit powers 
to reach a target signal-to-noise ratio (SNR), i.e., 
$\frac{|h_k|^2 P_k}{N_0} \ge \Gamma$,
where $\Gamma$ is the target SNR.
It is expected that $\bs_m$ is a $k_m$-sparse vector
(note that $k_m$ is the number 
of the active users in the $m$th channel).
Then, when $L > T_{\rm p}$, using compressive sensing algorithms 
\cite{Eldar12},
the BS is able to estimate $\bs_m$ under certain conditions
(of $\bC$ and the maximum sparsity of
$\bs_m$) \cite{Donoho06} \cite{Candes06},
which has been discussed in the context of compressive
random access.
Once $\bs_m$ is estimated, 
the determination of the number of active users 
or the estimation of $k_m$ becomes 
straightforward (because $k_m$ can be found from the sparsity
of $\bs_m$),
while preamble collision\footnote{It can happen
as a preamble can be chosen by multiple active users.}
\cite{Choi18c} \cite{Seo19}
can result in errors in estimating $k_m$.
From \cite{Mitz05},
for the $m$th channel, 
the conditional probability of no preamble collision
can be found as
\begin{align}
\uP_m (k_m)
& = \prod_{k=1}^{k_m-1} \left(1 - \frac{k}{L} \right) 
\approx e^{- \frac{k_m (k_m-1)}{2 L}},
\end{align}
where the approximation is actually a lower-bound
(thus, $1 - e^{- \frac{k_m (k_m-1)}{2 L}}$ is an upper-bound
on the conditional probability of preamble collision).
If $K$ active users are uniformly distributed
over $M$ channels and $K$ is assumed to follow a Poisson 
distribution with mean $\lambda$,
it can be shown that
\begin{align}
\uP_m 
& = \sum_{k_m = 0}^\infty e^{- \frac{k_m(k_m-1)}{2 L}}
p_{\frac{\lambda}{M}} (k_m) \cr
& \approx
\sum_{k_m = 0}^\infty 
\left(1- \frac{k_m(k_m-1)}{2 L} \right) p_{\frac{\lambda}{M}} (k_m) 
= 1 - \frac{\lambda^2}{2 L M^2}, \quad
\end{align}
where $\frac{\lambda}{M} < 1$.
Thus, $\frac{\lambda^2}{2 L M^2}$ 
becomes the probability of preamble collision.
It can be shown that
\be
1 - \uP_m 
\le \delta \Rightarrow  \frac{\lambda^2}{2 L M^2} \le \delta,
\ee
where $\delta \ll 1$ is a threshold probability of preamble collision.
To keep $\delta$, we need
\be
L \ge \frac{\lambda^2}{2 \delta M^2}.
	\label{EQ:Lge}
\ee

\begin{figure}[thb]
\begin{center}
\includegraphics[width=\figwidth]{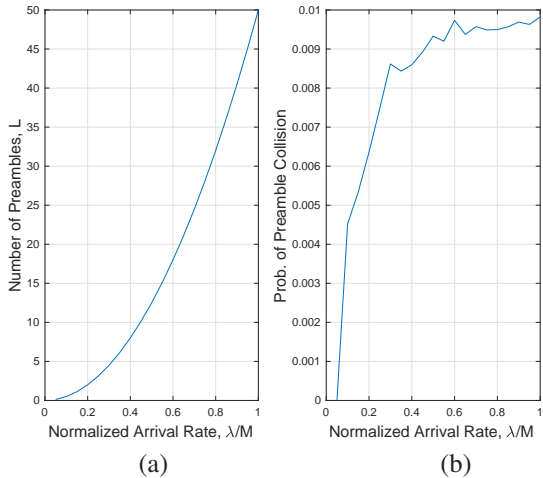} \\
\hskip 0.8cm (a) \hskip 3.5cm (b) \\
\end{center}
\caption{The number of
preambles and probability of preamble collision:
(a) $L = \frac{\lambda^2}{2 \delta M^2}$ 
versus $\frac{\lambda}{M}$
for a probability of preamble
collision of $\delta = 0.01$; (b) the probability of preamble collision 
versus $\frac{\lambda}{M}$
with $L = \lceil \frac{\lambda^2}{2 \delta M^2} \rceil$.}
        \label{Fig:Lkm}
\end{figure}

For example, letting $\delta = 0.01$ (i.e., the probability of 
preamble collision is less than 0.01),
using \eqref{EQ:Lge},
the number of preambles to keep 
the probability of preamble collision lower than $0.01$
is shown in Fig.~\ref{Fig:Lkm} (a) and 
the actual probability of preamble collision is shown 
in Fig.~\ref{Fig:Lkm} (b)
with $L = \lceil \frac{\lambda^2}{2 \delta M^2} \rceil$.

In the second case, it is assumed that all users
have unique preamble sequences. In this case, there is no preamble collision.
However, there are
a large number of columns of the matrix of preambles
(which is the same as the number of all users),
which makes the sparse signal estimation (and
the determination of the number of active users) difficult.
To avoid it, as suggested in \cite{Choi_18Feb},
sparse preamble sequences can be used.

In summary, by exploiting the notion of compressive sensing,
it is possible to determine $k_m$ at the BS. 
Note that, in the first case,
it is not necessary that the preamble sequences 
are orthogonal. For example, we can use
Zadoff-Chu or Alltop sequences \cite{Foucart13} for
preambles with reasonably low cross-correlation. 
In this 
case, the number of
preambles becomes $L = T_{\rm p}^2$ (for Alltop sequences)
when $T_{\rm p} \ge 5$ is a prime.
That is, a large number of preambles
to keep the probability of preamble collision 
low can be obtained with a reasonable length of preamble, $T_{\rm p}$.

\subsection{Downlink for Feedback}

As mentioned earlier, in multichannel ALOHA with EP,
the BS needs to feed back the number of active users
in each channel, $\{k_1, \ldots, k_M\}$. Thus,
if $n_{\rm f}$ bits are allocated for each $k_m$,
there might be $n_{\rm f} M$ bits required for the feedback.
Here, $n_{\rm f} = \lceil \log_2 \max k_m \rceil$, where 
$\max k_m$ might be a constant.

In fact,
the number of feedback bits can be reduced.
For each channel, it is necessary to send one bit:
$b_m = 1$ if $k_m = 1$ and $b_m = 0$ otherwise,
where $b_m$ is one-bit feedback for channel $m$.
If an active user that transmits a preamble to channel $m$ 
receives $b_m = 1$, this user belongs to Group I (i.e., contention-free),
and $\cS  = \{m\,|\, b_m = 1\}$.
Otherwise, an active user becomes a member of Group II.
In this case, to decide $p_{\rm dtp}$,
the BS needs to send additional information, which is $W$,
while $L$ can be found at any active user from $\{b_m\}$
as $L = M - \sum_{m=1}^M b_m$.
Thus, a total number of feedback bits is $M + 
\lceil \log_2 \max W \rceil$.

\section{Simulation Results}
\label{S:Sim}

In this section, we present simulation results 
for multichannel ALOHA when $K$ follows a Poisson distribution
with mean $\lambda$.
In addition, we only consider the case that 
$\lambda \le M$. Note that if $\lambda > M$,
the system is overloaded. In  
this case, since there might be more active users than
channels, it is expected that $k_m > 1$ for most $m$.
Therefore, the exploration gain would be diminished
and multichannel ALOHA with EP becomes less useful.

\begin{figure}[thb]
\begin{center}
\includegraphics[width=\figwidth]{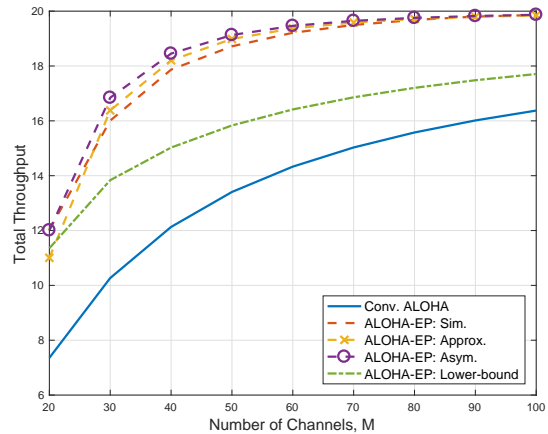}
\end{center}
\caption{Total throughputs 
of conventional multichannel ALOHA and
multichannel ALOHA with EP as functions
of $M$ when $\lambda = 20$.}
        \label{Fig:Tplt2}
\end{figure}

Fig.~\ref{Fig:Tplt2} shows the total throughputs 
of
conventional multichannel ALOHA and
multichannel ALOHA with EP as functions
of $M$ when $\lambda = 20$. We can see that the exploration
can help improve the throughput of multichannel ALOHA.

\begin{figure}[thb]
\begin{center}
\includegraphics[width=\figwidth]{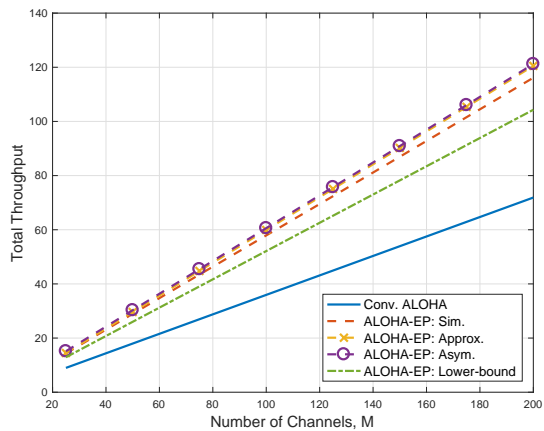}
\end{center}
\caption{Total throughputs 
of conventional multichannel ALOHA and
multichannel ALOHA with EP as functions
of $M$ when $\alpha = \frac{\lambda}{M} = 0.8$.}
        \label{Fig:Tplt3}
\end{figure}

In Fig.~\ref{Fig:Tplt3}, the total throughputs 
of conventional multichannel ALOHA and
multichannel ALOHA with EP are shown as functions
of $M$ when $\alpha = \frac{\lambda}{M} = 0.8$ is fixed. 
Note that the difference between the throughputs grows
linearly with $M$. 

\section{Concluding Remarks}
\label{S:Conc}

In this paper, an exploration 
approach has been proposed for multichannel ALOHA by sending preambles
prior to packet transmissions to allow
active users to learn the state of contention.
We found that the exploration gain is 
$2 - e^{-1}$ in terms of the ratio 
of the maximum throughput
of multichannel ALOHA with EP to that of 
conventional multichannel ALOHA.
Thanks to the improved throughput,
the proposed multichannel ALOHA scheme with EP 
becomes suitable for massive MTC.

\section*{Acknowledgement}

This research was supported
by the Australian Government through the Australian Research
Council's Discovery Projects funding scheme (DP200100391).

\bibliographystyle{ieeetr}
\bibliography{mtc}

\end{document}